\documentclass[sigconf]{acmart}

\usepackage{booktabs} 
\usepackage{graphicx}
\usepackage{subfigure}

\copyrightyear{2019}
\acmYear{2019} 
\setcopyright{iw3c2w3}
\acmConference[WWW '19]{Proceedings of the 2019 World Wide Web Conference}{May 13--17, 2019}{San Francisco, CA, USA}
\acmBooktitle{Proceedings of the 2019 World Wide Web Conference (WWW '19), May 13--17, 2019, San Francisco, CA, USA}
\acmPrice{}
\acmDOI{10.1145/3308558.3313527}
\acmISBN{978-1-4503-6674-8/19/05}

\usepackage{balance}

\begin{document}

\title{Judging a Book by Its Cover:\\ The Effect of Facial Perception on Centrality in Social Networks}

\author{Dongyu Zhang$^\ast$, Teng Guo$^\ast$, Hanxiao Pan$^\ast$, Jie Hou$^\ast$, Zhitao Feng$^\ast$, Yang Liang$^\dagger$, Hongfei Lin$^\dagger$, Feng Xia$^\ast$}

\affiliation{%
  {$^\ast$Key Laboratory for Ubiquitous Network and Service Software of Liaoning Province, School of Software,\\ Dalian University of Technology, Dalian 116620, China}
}
\affiliation{%
  {$^\dagger$School of Computer Science and Technology, Dalian University of Technology, Dalian 116024, China}
}
\email{%
  f.xia@acm.org
}

\begin{abstract}
	Facial appearance matters in social networks. Individuals frequently make trait judgments from facial clues. Although these face-based impressions lack the evidence to determine validity, they are of vital importance, because they may relate to human network-based social behavior, such as seeking certain individuals for help, advice, dating, and cooperation, and thus they may relate to centrality in social networks. However, little to no work has investigated the apparent facial traits that influence network centrality, despite the large amount of research on attributions of the central position including personality and behavior. In this paper, we examine whether perceived traits based on facial appearance affect network centrality by exploring the initial stage of social network formation in a first-year college residential area. We took face photos of participants who are freshmen living in the same residential area, and we asked them to nominate community members linking to different networks. We then collected facial perception data by requiring other participants to rate facial images for three main attributions: dominance, trustworthiness, and attractiveness. Meanwhile, we proposed a framework to discover how facial appearance affects social networks. Our results revealed that perceived facial traits were correlated with the network centrality and that they were indicative to predict the centrality of people in different networks. Our findings provide psychological evidence regarding the interaction between faces and network centrality. Our findings also offer insights in to a combination of psychological and social network techniques, and they highlight the function of facial bias in cuing and signaling social traits. To the best of our knowledge, we are the first to explore the influence of facial perception on centrality in social networks.
\end{abstract}

\begin{CCSXML}
	<ccs2012>
	<concept>
	<concept_id>10010405.10010455.10010459</concept_id>
	<concept_desc>Applied computing~Psychology</concept_desc>
	<concept_significance>500</concept_significance>
	</concept>
	<concept>
	<concept_id>10003033.10003083.10003090.10003091</concept_id>
	<concept_desc>Networks~Topology analysis and generation</concept_desc>
	<concept_significance>500</concept_significance>
	</concept>
	<concept>
	<concept_id>10010405.10010455.10010461</concept_id>
	<concept_desc>Applied computing~Sociology</concept_desc>
	<concept_significance>500</concept_significance>
	</concept>
	</ccs2012>
\end{CCSXML}

\ccsdesc[500]{Applied computing~Psychology}
\ccsdesc[500]{Applied computing~Sociology}
\ccsdesc[500]{Networks~Topology analysis and generation}

\keywords{Social Networks; Facial Perception; Bias; Centrality; Psychological Traits}

\maketitle

\section{INTRODUCTION}
Many cultures share beliefs about the human face being a window to a person’s characters. Humans frequently make trait judgments from facial clues \cite{todorov2009evaluating,willis2006first,zebrowitz1996wide}. Studies have indicated that 100 milliseconds of exposure to unfamiliar faces is sufficient for participants to form impressions about competence or trustworthiness \cite{willis2006first}.  Indeed, some have reported that people can make social judgments with exposures of less than 100 milliseconds, with any longer time being used to confirm judgments \cite{borkenau2009extraversion,porter2008face,todorov2009evaluating}. Moreover, there is a consensus concerning trait judgments based on faces among adults and even among 3- to 4-year-old children  \cite{cogsdill2014inferring}, despite the limited evidence of the validity of these judgments \cite{rule2013accuracy}.

Although these facial impressions of traits may not be accurate, they are very important within social network environments, because they may influence social behaviors such as seeking certain individuals for help, advice, dating, and cooperation \cite{stirrat2010valid,verplaetse2007you}. For example, people invest more money with those who they perceive as more trustworthy in economic games \cite{ewing2015perceived,rezlescu2012unfakeable}. Evaluations of faces have impacted decisions in electoral politics \cite{todorov2005inferences},  mate preferences \cite{little2006good}, hypothetical crime verdicts \cite{porter2010dangerous}, as well as approach or avoidance behaviors \cite{wilson2015facial, kong2019human}  as well as approach or avoidance behaviours \cite{todorov2008evaluating}.  In particular, regarding network centrality, individuals’ roles and relationships with others are more likely to be influenced by facial appearance at the early stage of central networks construction without or with little interaction or information on others.

However, little to no research has explored the effect of facial perceptions on social networks. Although previous work has investigated the relationships between trait perceptions and social networks, those perceptions are based on assessments of personality, behavior, emotion, ability, etc. (e.g. \cite{Brass1993Potential,chiu2017managers,Feiler2015Popularity,liu2018artificial}) rather than appearances or facial perception. Moreover, recent work suggests that psychological traits (based on psychological scales) correlate with the centrality of social networks \cite{morelli2017empathy}. 
However, whether face-based perceived psychological traits relate to centrality of networks is scarcely known. We therefore aim to explore the relationship between facial appearance and social networks by exploring newly formed social networks in a first-year college residential area. We took face photos of participants and asked them to nominate community members associated with different networks. We then collected image evaluation data by asking other  participants to rate face photos for three main attributions: dominance, trustworthiness, and attractiveness on a seven-point Likert-type scale. Our results revealed different social networks in different freshman residential areas. Moreover, we found that perceived facial dominance, trustworthiness, and attractiveness are effective at predicting the centrality of people on different networks. Our contributions are as follows.

\begin{itemize}	
	\item We explore the use of perceived facial features in a social network environment, and particularly for network centrality, to discover whether facial appearance affects social networks, which has scarcely been done before.
	\item We investigate the predictability of network centrality by rating face stimuli. The experimental results indicate that face-based traits predict membership of freshman community networks at the initial stage of formation.
	\item We propose a novel framework to conduct social network analysis in a psychological setting, and particularly for corroboration of facial bias. The results contribute to psychological knowledge of facial perception, because they are consistent with previous research that facial appearance affects network-based social behaviors.
\end{itemize}

This paper is organized as follows. In next section, we discuss the related work. In section 3, we introduce the details of data collection. In section 4, we describe methods used in this study. In section 5, we show the experiment results. In final section, we present conclusion of this paper.

\section{RELATED WORK}
\subsection{Dimensions of Face Evaluation}
Oosterhof and Todorov \cite{oosterhof2008functional} provided empirical evidence contributing to our knowledge of face evaluation. To identify the dimensions on which individuals evaluate faces, they conducted a principal component analysis on various trait adjectives inferred from neutral faces, and they reduced them to two main axes; dominance and trustworthiness. They constructed computer models of three-dimensional face shapes that correlated with ratings of trustworthiness and dominance \cite{oosterhof2008functional}.They found that by moving from the negative to positive extremes (low and high ratings) of the trustworthiness dimension, the faces changed from expressing slight anger to slight happiness. Likewise, in the dominance dimension, the faces varied from expressing feminine and baby-faced traits to masculine and mature-faced traits. The trustworthiness dimension arguably signaled an individual’s harmful intent, whereas the dominance dimension signaled the capability to carry out the harmful intention \cite{oosterhof2008functional}. Hence, they suggested that perhaps these evaluations of emotional valence and dominance are rooted in evolutionary mechanisms of threat detection \cite{oosterhof2008functional}. 

Additionally, Sutherland et al \cite{sutherland2013social}. described a third dimension of youthful attractiveness. In their study, they created average faces from photographs of people perceived as high or low in intelligence, trustworthiness, dominance, and confidence, and participants rated them on various traits. The stimuli of faces did not control for variables like hairstyle and piercings \cite{sutherland2013social}. The stimuli of faces did not control for variables like hairstyle and piercings \cite{sutherland2013social}. Their results supported the two dimensions proposed by Oosterhof and Todorov \cite{oosterhof2008functional}, and they indicated a third dimension of facial youthful attractiveness, which Sutherland et al. argued was important in making facial evaluations \cite{sutherland2013social}. This additional dimension may be due to the wider age range of faces used in the study \cite{sutherland2013social}. The aforementioned studies suggest that the main attributions made to faces are dominance, trustworthiness, and attractiveness; thus, we conducted studies based on these three dimensions.

\begin{figure*}
	\includegraphics[height=7cm]{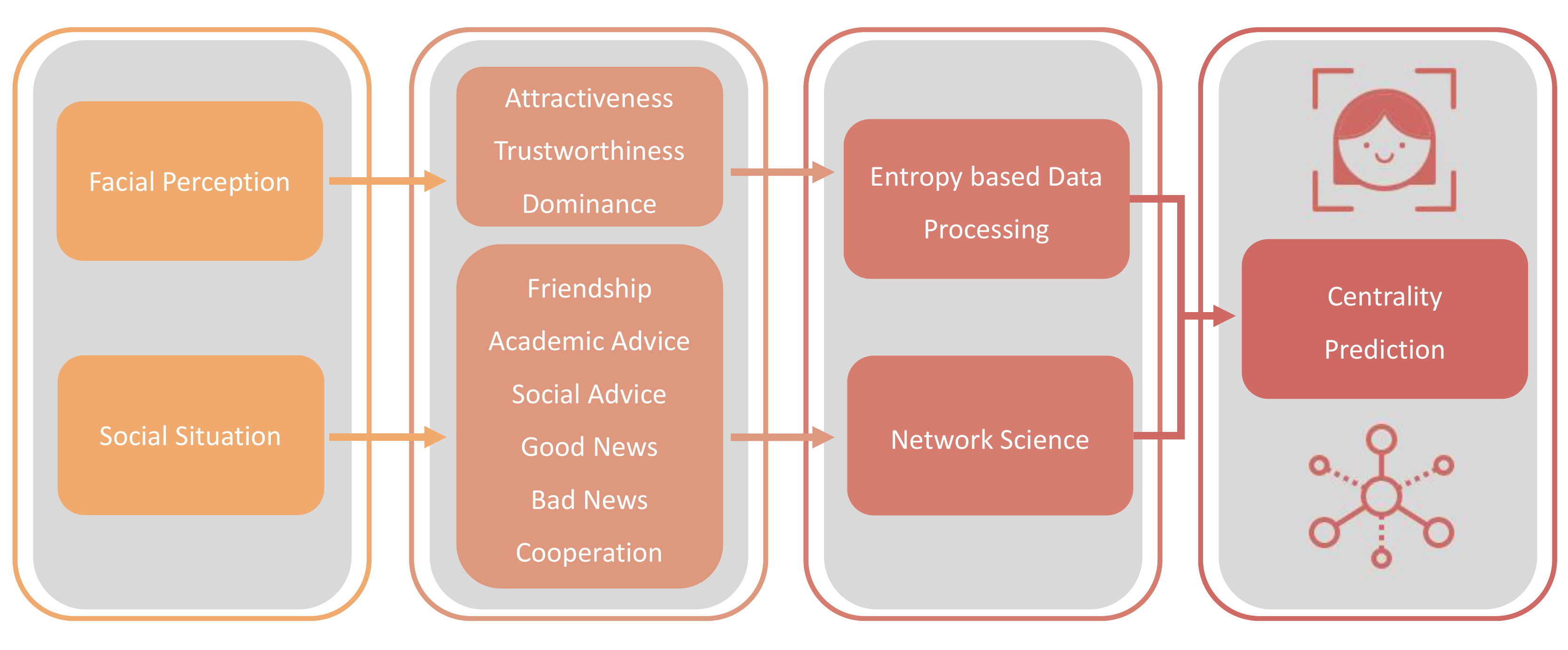}
	\caption{Methodology (PSP framework) used to analyze correlations between facial perception and social situation, and then to predict the centrality of different social networks. First, data of facial perception and social situation are collected from participants. Second, entropy-based data processing method and network science are used to process data. Finally, classification algorithm is used to predict the centrality in social networks according to facial perception features. }
	\label{intro}
\end{figure*}

Some studies have been done on how attractiveness, trustworthiness, and dominance impact network-related social behaviors. Alison et al. have studied how facial appearance can impact entrepreneurs to finance projects \cite{brooks2014investors}.
When entrepreneurs start a business, investors prefer pitches from attractive people, because they are much more persuasive. In strategic economic games, it is almost impossible for a player with an untrustworthy-looking face to be trusted by other players, even when he/she has a good credit record \cite{tingley2014face,rezlescu2012unfakeable}.
Moreover, Olivola et al. suggested that perceived dominance is correlated with professional success within the military and politics, because people believe that the leaders of these fields tend to look masculine, mature, and “cold”  \cite{olivola2014many}.
Therefore, since perceived trustworthiness, attractiveness, and dominance relate to social behavior in social network settings, they may influence network centrality.

\subsection{Positions in Social Networks}
People’s positions in social networks are affected by many factors, such as behavior, personality, and emotion. Previous research indicates that behavioral tactics are associated with certain structural positions, such as indegree centrality \cite{Brass1993Potential}. A person who is considered assertive and rational, and who has upward appeal, can also use behavioral tactics. These people are always considered dependable and reliable. Therefore, people tend to ask these persons to help them when they are in trouble, or they share good news with them. Thus, people like to follow those who have social power, and to give them central positions in social networks. A person’s behavior is the reflection of his/her personality. Psychological research suggests that people’s personalities have impacts on their positions in social networks. Feiler et al. \cite{Feiler2015Popularity} found that extraverted people have more friends than introverted people, and that two extraverts are more likely to become friends because of the homophily effect. Therefore, extraverted people are much more likely to be involved in social networks, which places them in central positions. Not only do different personalities have impacts on people’s positions in social networks, but also emotions can change people’s centrality in networks. People’s emotions also relate to their personalities. Research indicates that people who have positive attitudes to life are attractive in social networks \cite{chiu2017managers}. This is because people with high life satisfaction find it easy to attract their peers’ attention and alliances \cite{Lyubomirsky2005The}. Those who have positive emotions like to experience positive emotions. Therefore, such persons tend to become extraverted. As we all know, outgoing people are much more active in social networks. They are likely to communicate with others and to make their positions more central.

\section{DATA COLLECTION}
\subsection{Privacy Protection}
Prior to participation, all participants provided a consent form and consented to take part in the study. They were debriefed after taking part in the experiments. Participants were informed that data gathered during this research would be coded and kept confidentially by the researcher, with only the researcher and supervisor having access; all information is stored with a code, and there is no immediately identifiable information. Part of the research involved taking photographic images. These images are kept secure and stored with no identifying factors, i.e., consent forms and questionnaires. 
\subsection{Facial Image and Social Network Data Collection}
We used both online and offline campus advertisements to recruit college freshman participants who might be interested in this research in the university’s freshman residential area. Student participants were required to be more than 18-year-old freshmen, who lived in several specific freshman residential buildings (next to each other) in the same area.  A total of 185 participants (aged 18-20, mean = 19.07, SD = 0.18, males=102, female=84) were recruited (89\% of the freshman in this area). They were called to the lab, they completed a social network questionnaire, and their photos were taken. This took place at the beginning of third month of freshman registration of university entrance, and it lasted for one month. Monetary compensation was given for participation in the study.

When participants arrived at the lab, prior to participation, they provided a consent form and consented to take part in the study. Participants first reported their age, gender, country of birth, country of residence, and ethnicity via a questionnaire. They then listed up to eight people on six questions in random order with reference to additional name lists, which gave the names of all freshmen living in this area. Eight people should be listed with the priority from high to low: “1” stands for the highest priority, and priority decreases in turn; “8” stands for the lowest priority. The six questions are as follows: “Please list your good friends.” “To whom will you turn for help, when you have difficulties in study?” “To whom will you turn for help, when you encounter some troubles in life?” “When you get a piece of good news, with whom will you share it?” “When you get a piece of bad news, with whom will you share it?” “Whom do you want to invite to join your team when you have a task that requires a team to accomplish it?”
We also took face photographs of students after they completed the network questionnaire. Participants were instructed to look at the center of the camera lens with neutral expressions, hair pulled back, and no adornment. They were photographed under standardized lighting conditions with a fixed camera distance. They were debriefed after taking part in the experiments. A set of 185 images of 185 identities in six groups (30 or 31 images of each group) were thus available for the study. 

\subsection{Facial Perception Data Collection}
A total of 180 participants (no freshmen)  were recruited in the university with payment(aged 18-27, mean = 23.09, SD = 3.06, males=112, female=68). We used different participants from those freshmen to rate images because we focus on the effect of  facial bias which frequently made by strangers. They were called to the lab for face rating by appointment. The set of stimuli for rating included 30 identity images (30 of 185 images in each block with no face identities repeated) with randomized allocation for different participants. Each image was rated by 30 participants. Each participant saw 30 images in a random order. Participants completed three ratings: trustworthiness, attractiveness, and dominance. Participants saw one image at a time. For trustworthiness (attractiveness or dominance) ratings, participants were instructed “You will make judgements regarding trustworthiness (attractiveness or dominance). Please rate your impression from score 1 to 7 on how trustworthy (attractive or dominant) the face appears”; they then rated the stimuli for perceived trustworthiness (attractiveness or dominance) on a seven-point Likert-type scale labelled (1) very negative and (7) very positive, with the question on the top of each facial image being “how trustworthy (attractive or dominant) does this face appear?” For attractiveness rating, participants were given the instruction “You will make judgements regarding attractiveness. Please rate your impression of how attractive this face is from 1 to 7.” Participants rated apparent attractiveness on this seven-point scale.

\section{METHODS}
To solve this kind of problem that explores the relation between facial perception and social network situation, we proposed a framework named PSP (Facial Perception and Social Situation Prediction Framework) (shown in Figure \ref{intro}).
In this framework, we proposed an entropy-based data processing method to remove the noise in questionnaire data. Each part included in the framework is introduced as follows.

\subsection{Problem Formulation}
We assumed that there is a photo of student $i$ and this photo is evaluated by thirty people from three aspects: attractiveness, trustworthiness and dominance. In other word, the photo of student $i$ will get three scores {[$s_{i1},s_{i2},s_{i3}$]}. In addition, we calculated the degree centrality of student $i$ in six different social networks{[$d_{i1},d_{i2},d_{i3},d_{i4}$,$d_{i5}$,
$d_{i6}$]}. Then, we compute the correlation between photo scores and degree centrality of each network and explore the predictability of degree centrality using the photo scores.

\subsection{Network Analysis}
We carried out a detailed analysis of six social networks through several parameters as follows, in order to get a deep understanding to sample's social situation.
\subsubsection{Degree correlation and assortativity coefficient}
The degree distribution, which is the probability that there are nodes in a network with the degree of $i$, is defined as
\begin{equation}
P(i) = \frac{n(i)}{N},
\end{equation}
where $n(i)$ is the number of nodes with degree $i$ and $N$ is the total numbers of nodes in the network.
The joint probability distribution of the network is defined as the proportion of edges between nodes with degree $i$ and $j$:
\begin{equation}
P(i, j) = \frac{m(i, j)\mu(i, j)}{2M},
\end{equation}
where $m(i, j)$ is the number of edges between nodes with degree $i$ and $j$ and $M$ represents the total number of edges. If $i = j$, then $\mu(i, j) = 2$, otherwise $\mu(i, j) = 1$. Then, the excess degree distribution is defined as 
\begin{equation}
P_n(i) = \sum_{i = j_{min}}^{j_{max}}P(i, j),
\end{equation}
where $j_{min}$ and $j_{max}$ are the minimum and maximum of nodes' degree in the network. The excess degree distribution indicates the probability that of choosing a node with degree $i$ from the neighbors of a node chosen randomly.

Denote the degree distribution $P(i)$ as $p_i$, the joint probability distribution $P(i, j)$ as $e_{ij}$, and the excess degree distribution $P_n(i)$ as $q_i$, then the assortativity coefficient is defined as
\begin{equation}
r = \frac{1}{\sigma_q^2}\sum_{i, j}ij(e_{ij} - q_i q_j),
\end{equation}
where $\sigma_q^2 = \sum_{i}i^2 q_i^2 - [\sum_i i q_i]^2$ is the variance of the excess degree distribution $q_i$. The assortativity coefficient $r$ ranges from $-1$ to $1$. If $r > 0$, then the network is assortative, which means that nodes with bigger degree prefer to connect with nodes with bigger degree, and vice versa. The absolute value of $r$ reflects the strength of assortativity or disassortativity of the network.

We can also use excess average degree to quantify the degree correlation of network approximately. The excess average degree neighbors is defined as
\begin{equation}
\langle k_{nn}\rangle_i = \frac{1}{k_i}\sum_{j = 1}^{k_i}k_{i_j},
\end{equation}
where ${k_i}k_{i_j}$ is the degree of number $i_j$th neighbor of node $i$.

\subsubsection{Clustering coefficient}
The clustering coefficient quantifies the probability that the neighbors of node $i$ are connected with each other. It is a frequently used metric for the community structure of a network. The clustering coefficient of a node $i$ with degree of $k_i$ is defined as
\begin{equation}
C_i = \frac{E_i}{(k_i(k_i - 1)) / 2} = \frac{2E_i}{k_i(k_i - 1)},
\end{equation}
where $E_i$ is the numbers of edges existing among the $k_i$ neighbors of node $i$.

\subsection{Correlation Coefficient}
Pearson’s correlation is used to explore the relation among facial perception features and spearman's correlation is used to explore the relation between facial perception feature and the centrality of six social networks because of the different scale.
\subsubsection{Spearman’s correlation}

Spearman’s correlation, also known as Spearman’s rank correlation, measures the correlation of rank between two populations. Here, we use it in the analysis of the relationships between the scores of photos and their centralities in different social networks. The Spearman’s rank correlation coefficient is defined as

\begin{equation}
r_S = 1 - \frac{6 \sum_{i = 1}^{N} d_{i}^{2}}{N(N^2 - 1)},
\end{equation}
where $N$ is the number of students in our experiment, and $d_i = r(S_i) - r(C_i)$, with $r(S_i)$ and $r(C_i)$ being the ranks of students' photo scores and centralities respectively.

\subsubsection{Pearson’s correlation}

Pearson’s correlation coefficient quantifies the linear correlation between two groups of continuous variables. The Pearson correlation coefficient is defined as

\begin{equation}
r_{X,Y} = \frac{E[(X - \mu_X)(Y - \mu_Y)]}{\sigma_X \sigma_Y},
\end{equation}
where $\mu$ and $\sigma$ are the mean value and standard deviation of random variable $X$ and $Y$, and $E(\cdot)$ is the expectation. Here, we use the Pearson correlation coefficient to analyze the linear relationships among three kinds of scores of students’ photos.

\subsection{Prediction}
First, we proposed a novel method that using information entropy to remove the noise in questionnaire data. Then, we redesigned our prediction task as binary classification task and used four classical classification algorithms to predict student centrality.
\subsubsection{Feature Processing}
As mentioned before, each photo was evaluated by a certain number of participants. Some participants evaluated all photos with same score or score with low separating capacity, because they didn’t take these questionnaires seriously. As the noise, these scores will impact experiment result. Here we borrowed the concept of entropy in information theory to remove these meaningless data. We proposed a participant entropy which is calculated according to the score they gave to each photo.
\begin{equation}
E = -\sum p(x)\log p(x)\
\end{equation}
where $p(x)$ represents the probability of occurrence of sample $x$.
\subsubsection{Prediction Models}
In the experiment, we used four classical classification methods to solve the prediction problem. First, we tried the k-nearest neighbor (KNN) classifier ~\cite{cattaneo2018large}, which finds the top k nearest neighbors of a test sample in the train set on the feature space and makes predictions by vote. The idea of this method is very naive, namely that if samples are closer in the feature space, they are more likely to be in the same class.

Then, we observed two linear classification models’ performance, linear support vector classifier (linear SVC)  ~\cite{bottou2018optimization, morente2017improving} and logistic regression (LR) ~\cite{chai2018novel}. Both identify hyperplanes in the feature space to split samples into different classes. However, these two algorithms obtain the best partition in different ways. The linear SVC looks for a hyperplane that maximizes the interval between two classes, that is,
\begin{equation}
\left\{
\begin{aligned}
max_{\textbf{w}, b} & \quad \frac{2}{\|\textbf{w}\|} & \\
s.t.&\quad y_i (\textbf{w}^T x_i) \geq 1, & i = 1, 2, 3\cdots , \\
\end{aligned}
\right.
\end{equation}
where $y_i$ and $x_i$ are labels and features of $i$th sample in the train set, and \textbf{w} is the normal vector of the hyperplane.

Logistic regression estimates the relative logarithm odds of a sample as a positive one, that is
\begin{equation}
\textbf{w}^T \textbf{x} + b = ln\frac{y}{1 - y},
\end{equation}
where $y$ is the odds of sample being a positive one, \textbf{x} is its feature vector, and \textbf{w} is its corresponding coefficient.

We also used a probability model, a Naive Bayesian classifier, which is based on the attribute conditional independence assumption,
\begin{equation}
P(c|\textbf{x}) = \frac{P(c)P(\textbf{x}|c)}{P(\textbf{x})} = \frac{P(c)}{P(\textbf{x})}\prod_{i = 1}^{d}P(\textbf{x}_i|c),
\end{equation}
where $P(c|\textbf{x})$ is the probability of a sample belonging to class c under the condition that its feature vector equals \textbf{x}. The Naive Bayesian classifier involves a prior assumption of the sample distribution. Here we used the Gaussian distribution due to the distribution of data.

\section{RESULTS}
\subsection{Facial Perception Analysis}

\begin{figure}
	\centering
	\includegraphics[height=8cm]{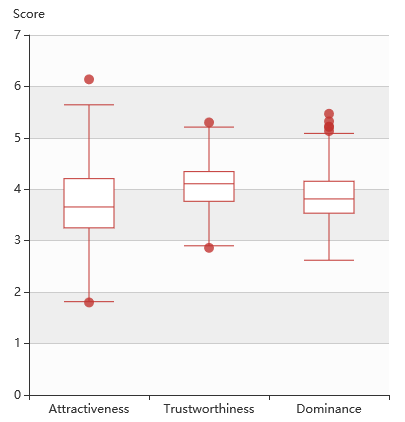}
	\caption{Box plot of attractiveness, trustworthiness, and dominance.}
	\label{boxplot}
\end{figure}

\begin{figure*}[t]
	\subfigure[Friendship]{
		\begin{minipage}{5.6cm}
			\centering
			\includegraphics[height=4.5cm]{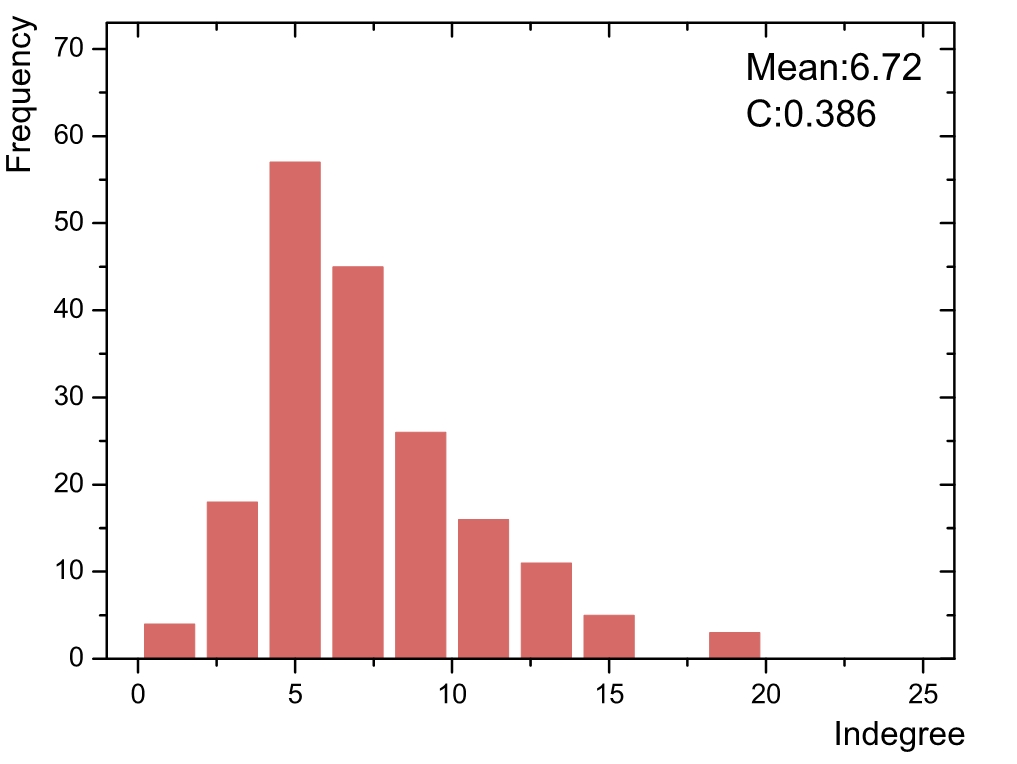}
		\end{minipage}
	}
	\subfigure[Academic Advice]{
		\begin{minipage}{5.6cm}
			\centering
			\includegraphics[height=4.5cm]{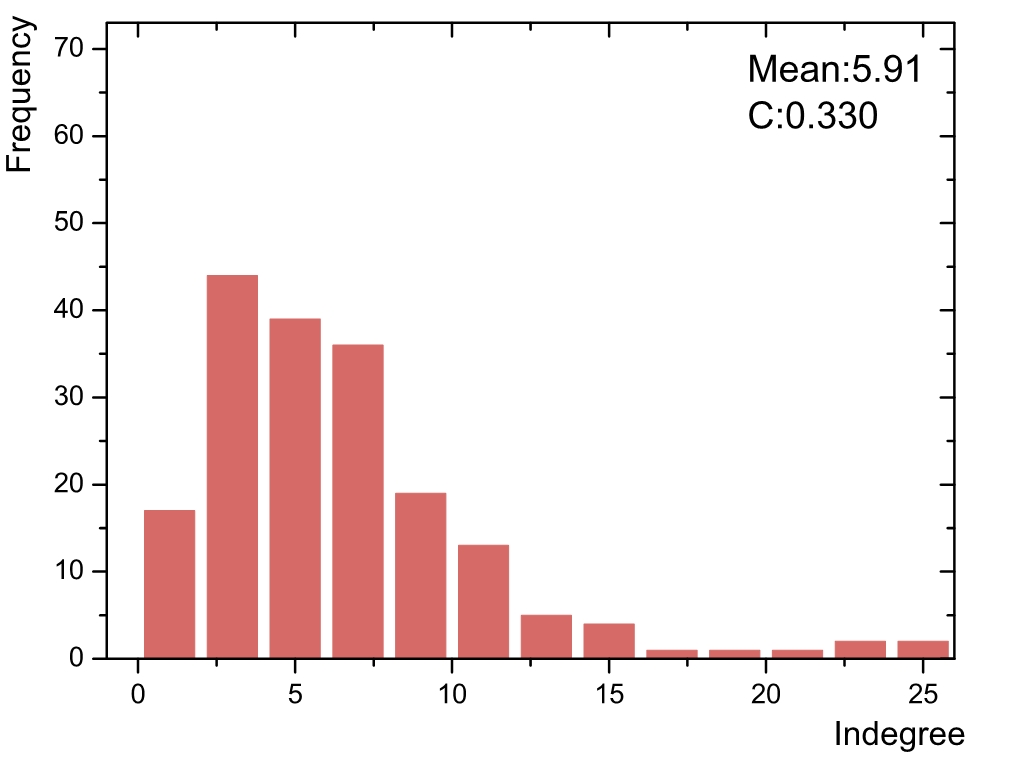}
		\end{minipage}
	}	
	\subfigure[Social Advice]{
		\begin{minipage}{5.6cm}
			\centering
			\includegraphics[height=4.5cm]{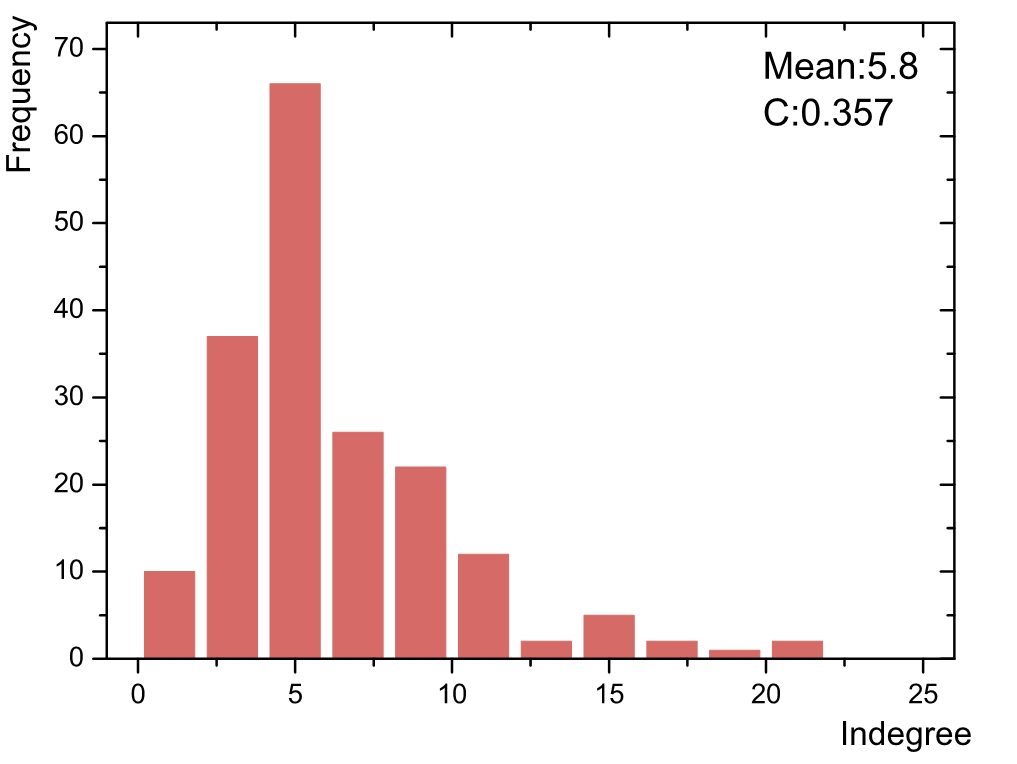}
		\end{minipage}
	}
	\subfigure[Good News]{
		\begin{minipage}{5.6cm}
			\centering
			\includegraphics[height=4.5cm]{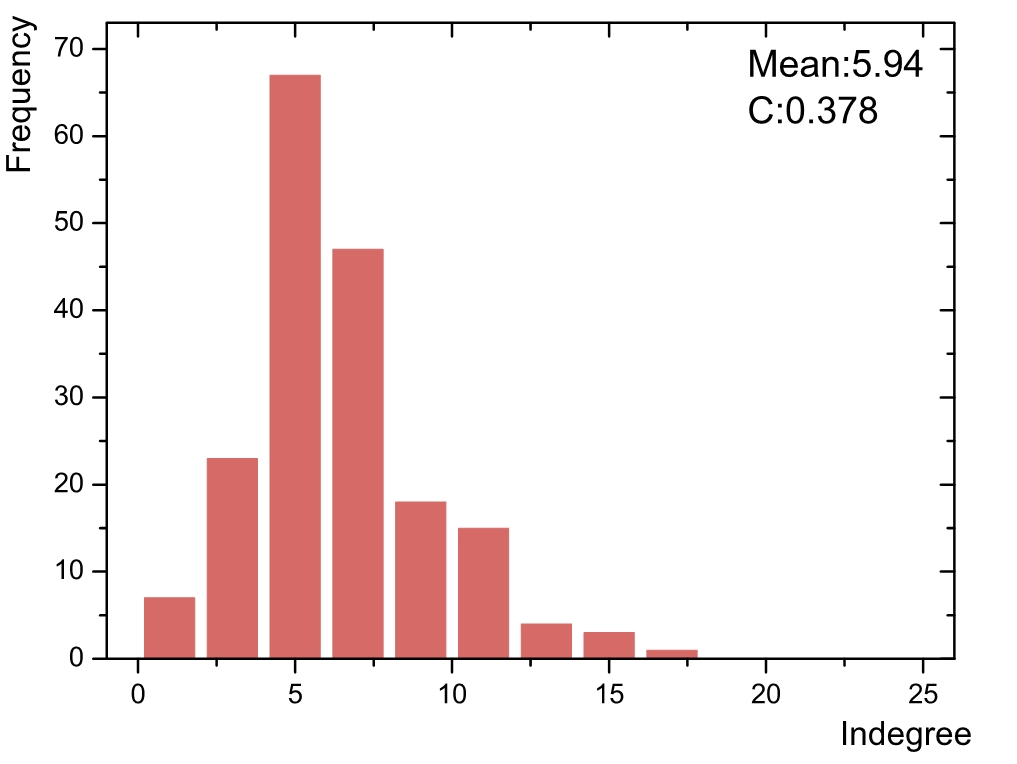}
		\end{minipage}
	}
	\subfigure[Bad News]{
		\begin{minipage}{5.6cm}
			\centering
			\includegraphics[height=4.5cm]{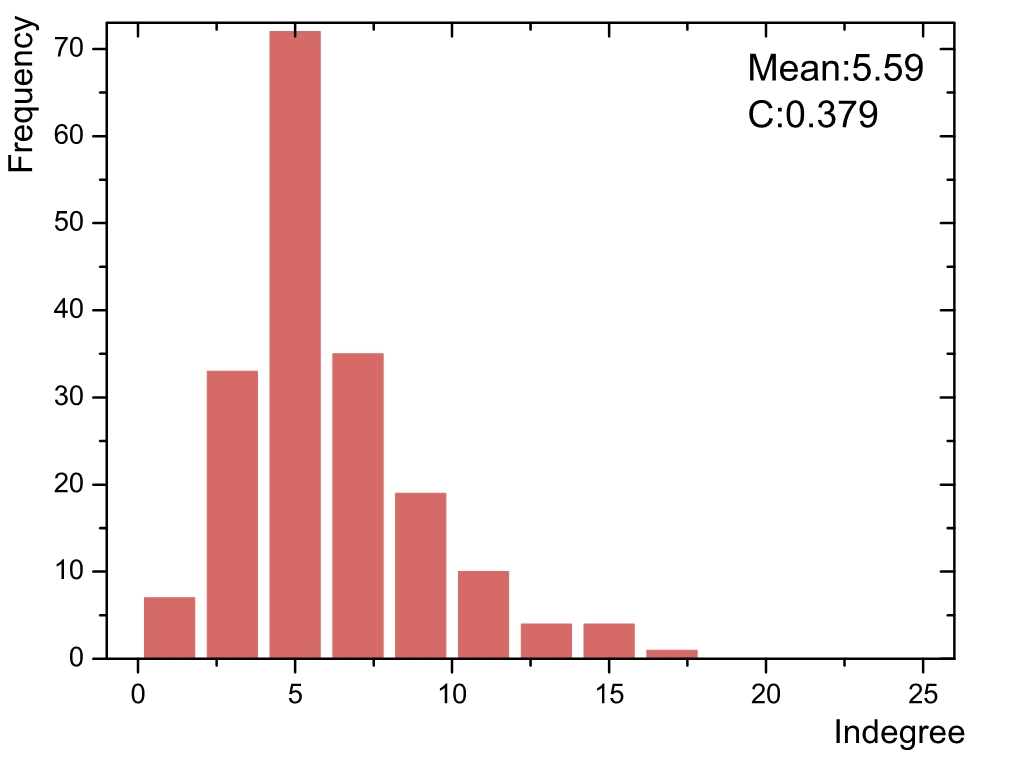}
		\end{minipage}
	}	
	\subfigure[Cooperation]{
		\begin{minipage}{5.6cm}
			\centering
			\includegraphics[height=4.5cm]{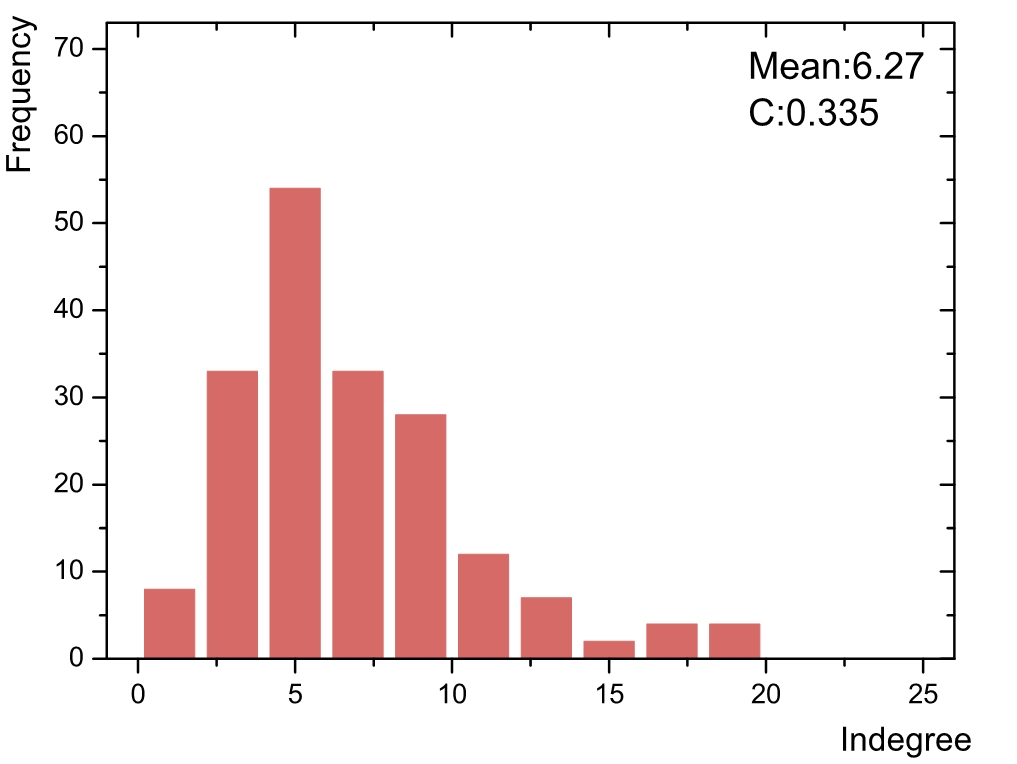}
		\end{minipage}
	}
	\caption{Histograms of indegree for the six social networks. In each figure, The 'Mean' and 'C' represent the average degree and clustering coefficient. }
	\label{fig:indegree} 
\end{figure*}
To explore the relationship between the centrality of social networks and perception, photos of participants were measured in three dimensions: attractiveness, trustworthiness, and dominance, and the distributions of data of each dimension are shown in Figure \ref{boxplot}. 
\begin{table}
	\caption{Pearson's correlation coefficient}
	\label{Pearson}
	
	\begin{tabular}{lccc}
		\hline
		Score           & Attractiveness & Trustworth- & Dominance \\ 
		&                &     iness    &             \\ \hline
		Attractiveness  & -              & -               & -         \\
		Trustworthiness & 0.410**        & -               & -         \\
		Dominance       & 0.379**        & 0.179*          & -         \\ \hline
	\end{tabular}
	\begin{flushleft}
		*$p$<0.05; **$p$<0.01  
	\end{flushleft}
	
\end{table}
The average score for trustworthiness was higher than the scores for other two aspects, and attractiveness scored lowest, which suggests that overall the freshmen prefer to trust strangers. For the dispersion level, the score for attractiveness was spread out further from its average value than others, as the span and interquartile range of attractiveness are much longer than the other two dimensions. For extreme outliers, compared with attractiveness and trustworthiness, participants give more extreme outliers on dominance. The distribution of data on these three dimensions has no significant skewness. In a word, Figure \ref{boxplot} suggests that students involved in our dataset covered almost all levels of facial appearance. Moreover, the correlations among attractiveness, trustworthiness, and dominance are shown in Table ~\ref{Pearson}. Attractiveness has a relatively strong positive correlation with trustworthiness and dominance. The correlation between trustworthiness and dominance is also positive, but relatively weak. This means that people with high attractiveness can be trusted easily, and they can also be deemed to dominate.

\begin{figure*}
	\subfigure[Friendship]{
		\begin{minipage}{5.6cm}
			\centering
			\includegraphics[height=4.5cm]{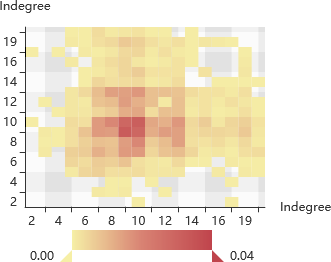}
		\end{minipage}
	}
	\subfigure[Academic Advice]{
		\begin{minipage}{5.6cm}
			\centering
			\includegraphics[height=4.5cm]{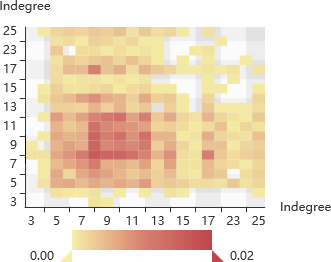}
		\end{minipage}
	}	\subfigure[Social Advice]{
		\begin{minipage}{5.6cm}
			\centering
			\includegraphics[height=4.5cm]{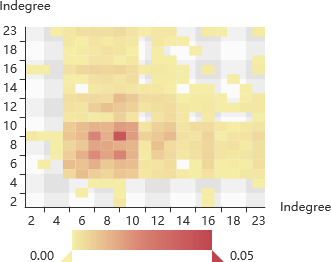}
		\end{minipage}
	}
	\subfigure[Good News]{
		\begin{minipage}{5.6cm}
			\centering
			\includegraphics[height=4.5cm]{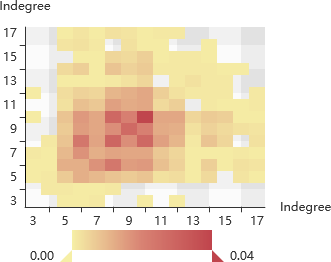}
		\end{minipage}
	}
	\subfigure[Bad News]{
		\begin{minipage}{5.6cm}
			\centering
			\includegraphics[height=4.5cm]{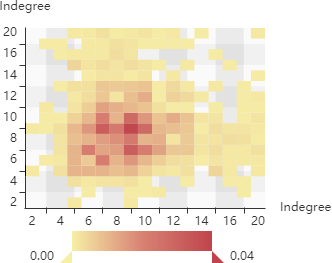}
		\end{minipage}
	}	\subfigure[Cooperation]{
		\begin{minipage}{5.6cm}
			\centering
			\includegraphics[height=4.5cm]{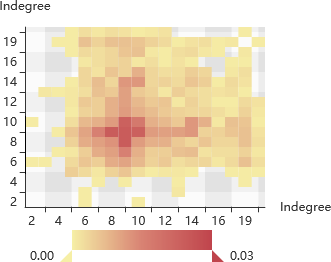}
		\end{minipage}
	}
	\caption{Joint probability distribution for the six social networks.}
	\label{Joint probability}
\end{figure*}

\subsection{Social Network Analysis}

We built six directed networks defined by different types of relationships from our questionnaire. The nodes are all connected in these six directed networks. As we can see from Figure ~\ref{fig:indegree}, the distributions of indegree for each of the six networks look like left-skewed bell-shaped curves. Due to the requirements of each questionnaire, most of the average indegree of the networks were about 6, except for the friendship network, with an average indegree of 6.72, which suggests that different friends meet different social requirement. For example, friends who are shopping together may not be suitable for starting a business together. However, in the bad news network, the average indegree is the lowest, and in most cases the indegree is around 5. This observation is consistent with general daily experience that bad news can only be shared with close friends. Moreover, about 75\% of the indegrees are between 3 and 8, and the most frequent indegree is 5. Only a few individuals have a big indegree, which means that others prefer to choose them as listeners or helpers. In other words, few individuals occupy a center position in the social networks. This verifies the view that the Matthew Effect exists in social networks. Furthermore, four networks’ clustering coefficients are around 0.37, while they are about 0.33 in the cooperation and academic advice networks. Compared to some typical social networks, all these clustering coefficients are lower. This is probably because the individuals we researched are freshmen, and their social networks are still at an early stage. Their circles of friend are still narrow, so some friends may not know each other. As time goes by, more commonality will develop between students, and increasing links will appear, resulting in a higher clustering coefficient.

\begin{figure*}
	\subfigure[Friendship]{
		\begin{minipage}{5.5cm}
			\centering
			\includegraphics[height=4.5cm]{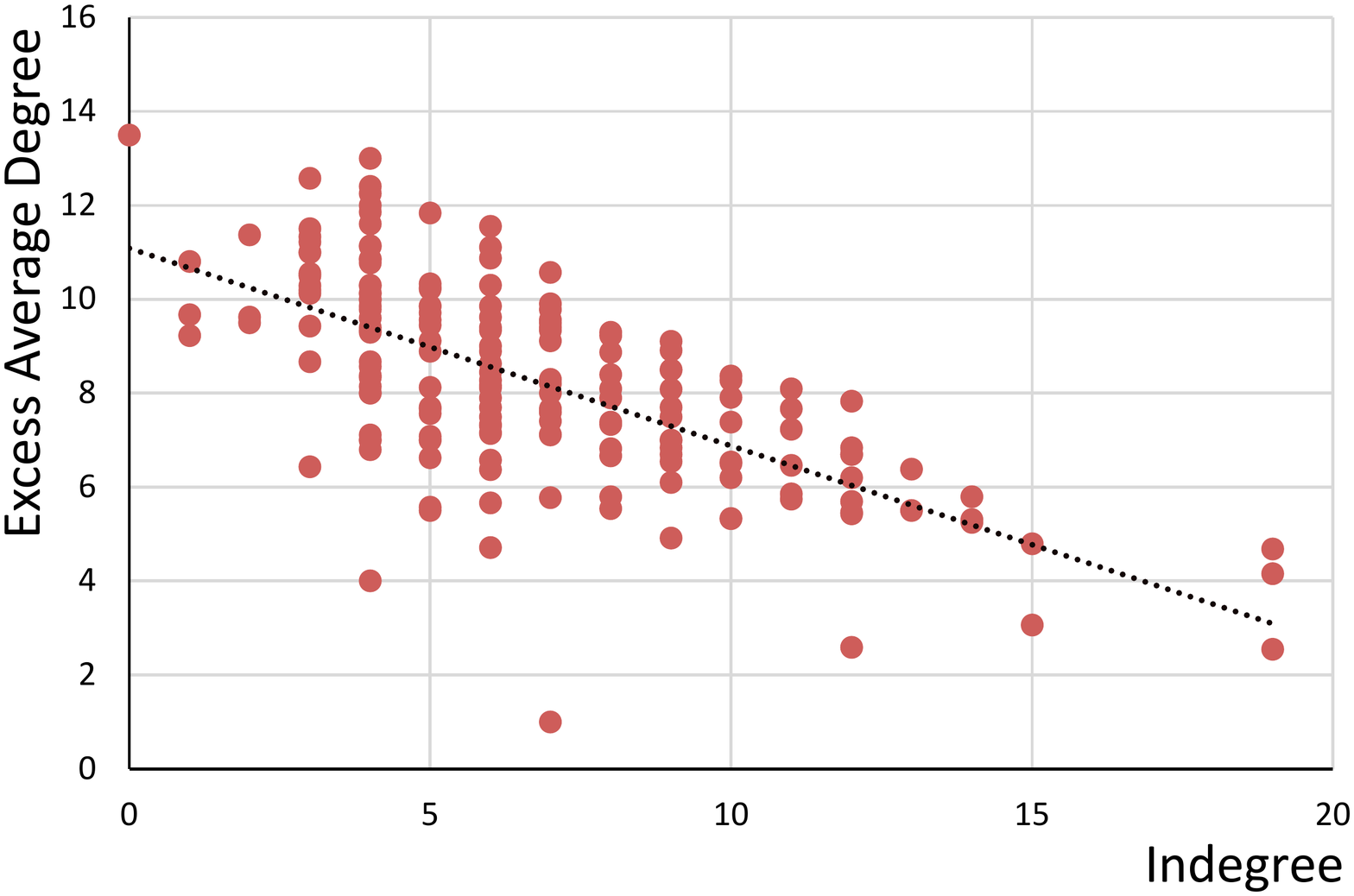}
		\end{minipage}
	}
	\subfigure[Academic Advice]{
		\begin{minipage}{5.5cm}
			\centering
			\includegraphics[height=4.5cm]{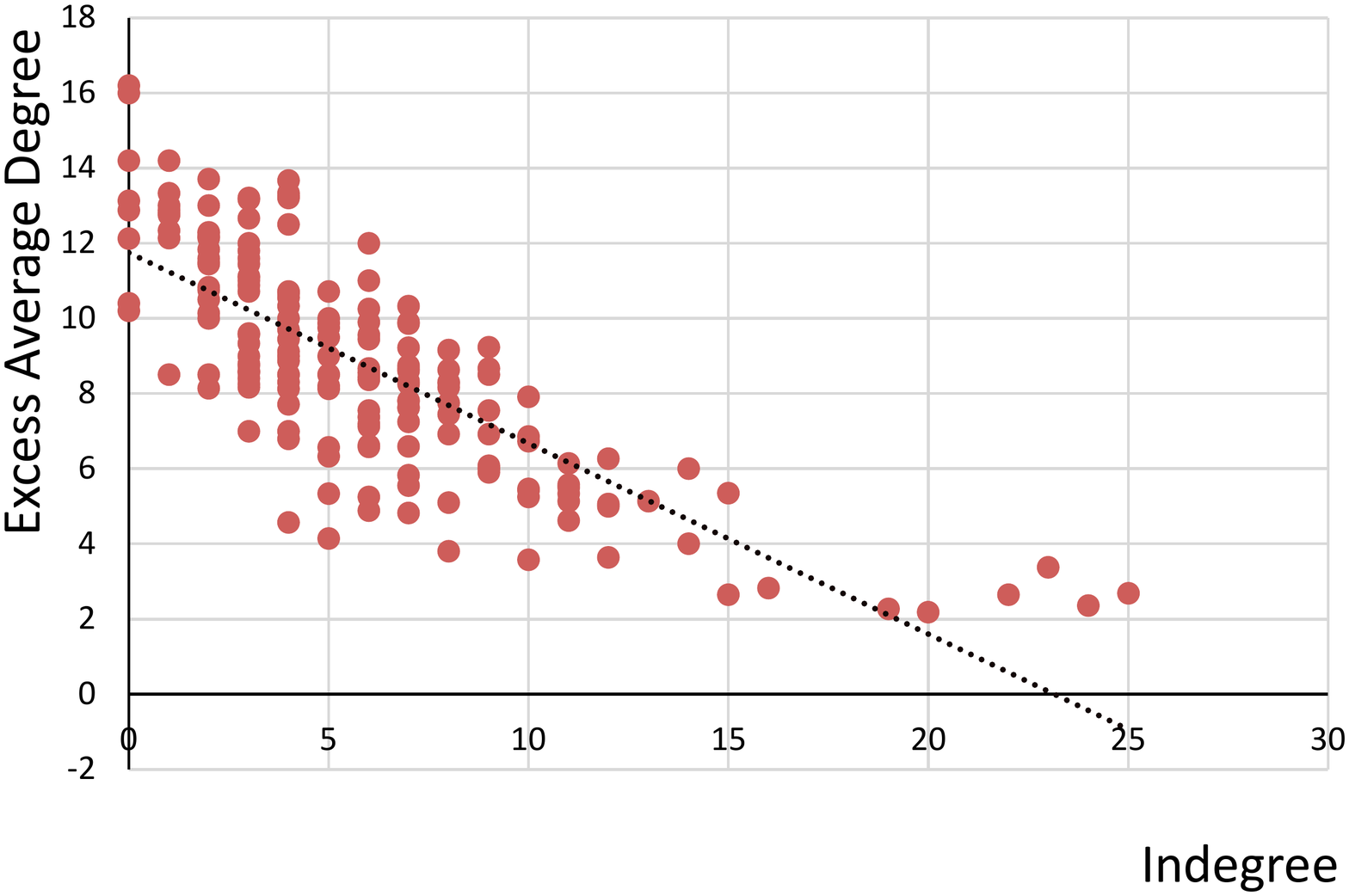}
		\end{minipage}
	}	\subfigure[Social Advice]{
		\begin{minipage}{5.5cm}
			\centering
			\includegraphics[height=4.5cm]{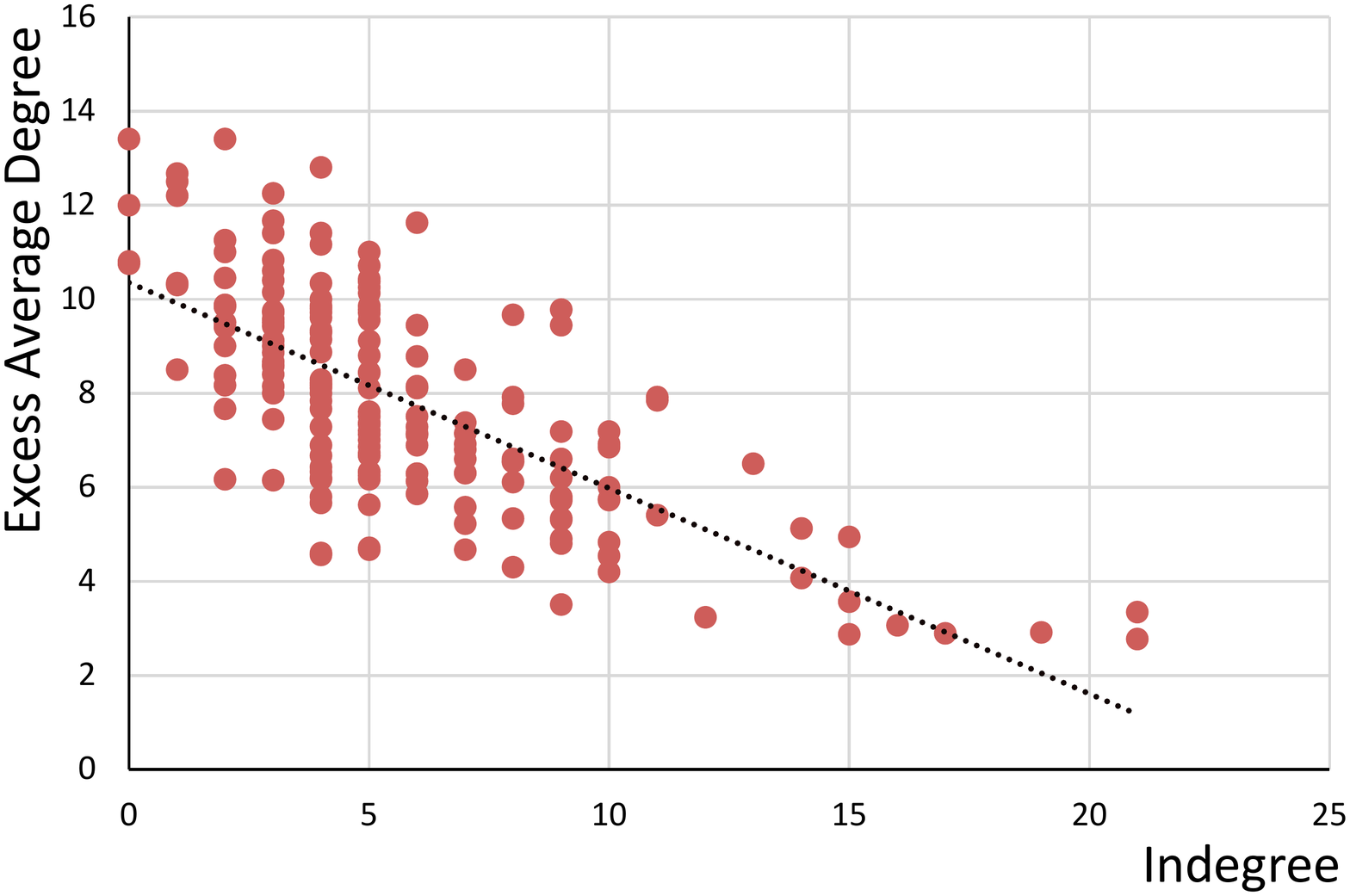}
		\end{minipage}
	}
	\subfigure[Good News]{
		\begin{minipage}{5.5cm}
			\centering
			\includegraphics[height=4.5cm]{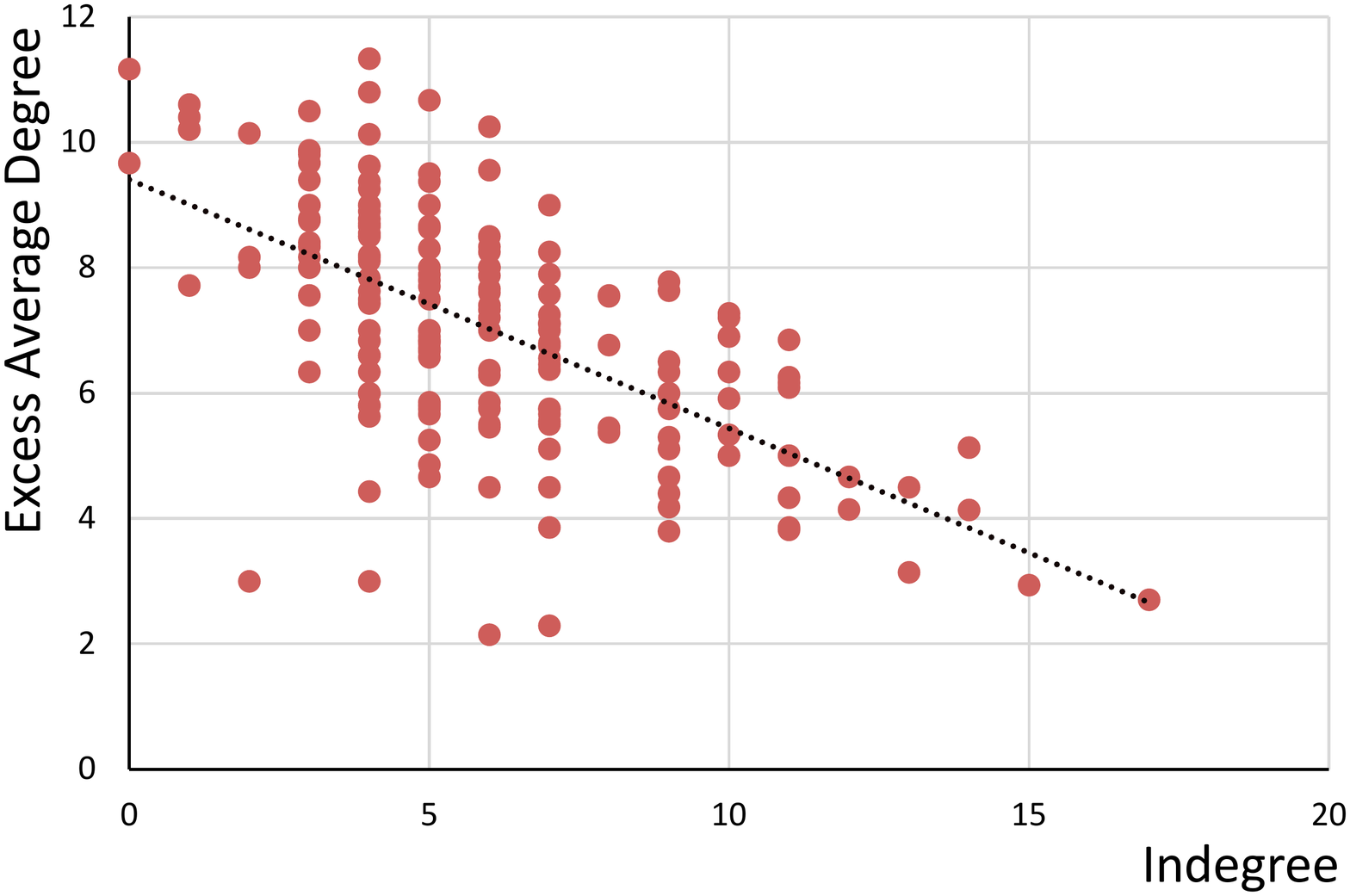}
		\end{minipage}
	}
	\subfigure[Bad News]{
		\begin{minipage}{5.5cm}
			\centering
			\includegraphics[height=4.5cm]{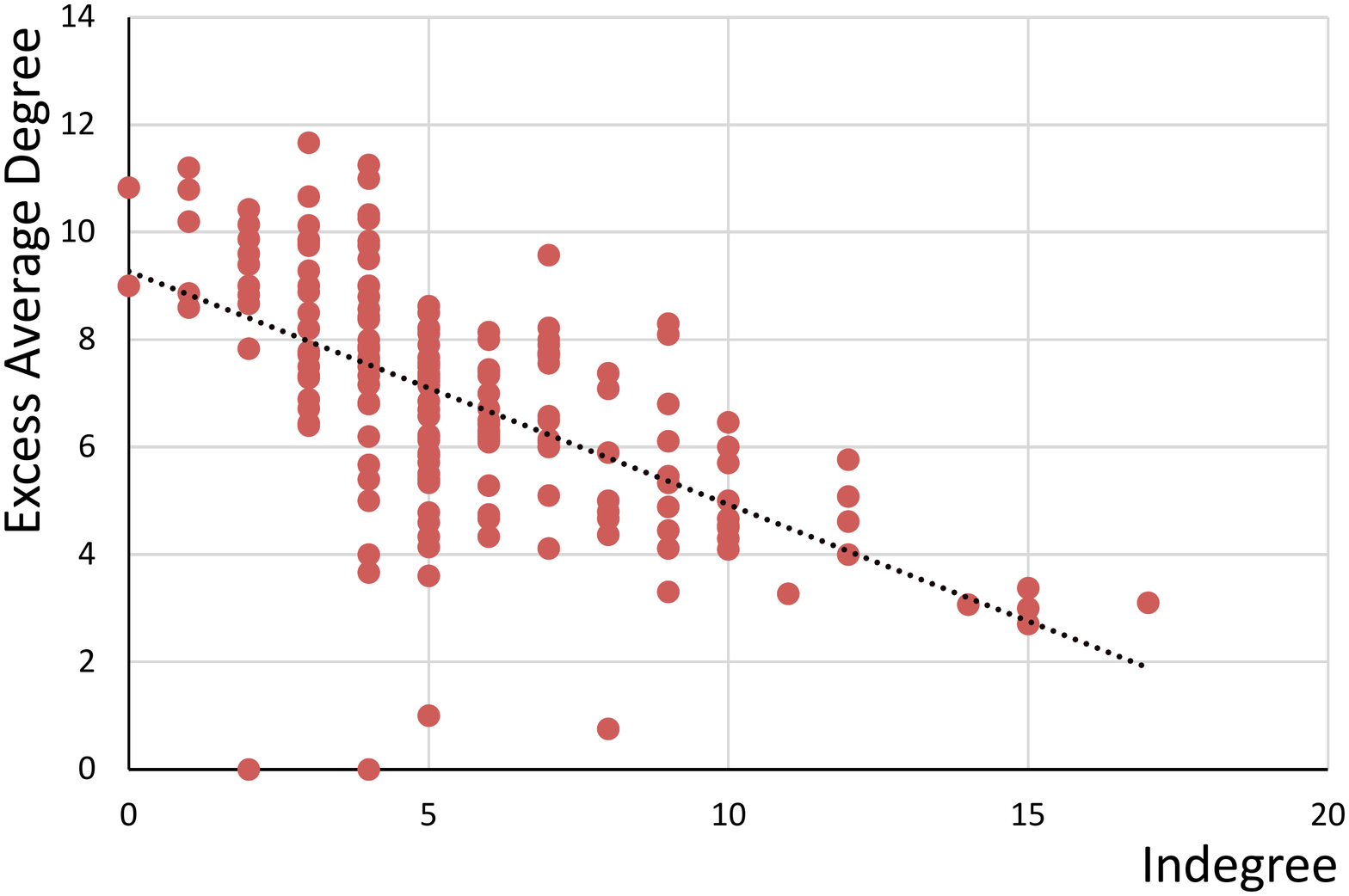}
		\end{minipage}
	}	\subfigure[Cooperation]{
		\begin{minipage}{5.5cm}
			\centering
			\includegraphics[height=4.5cm]{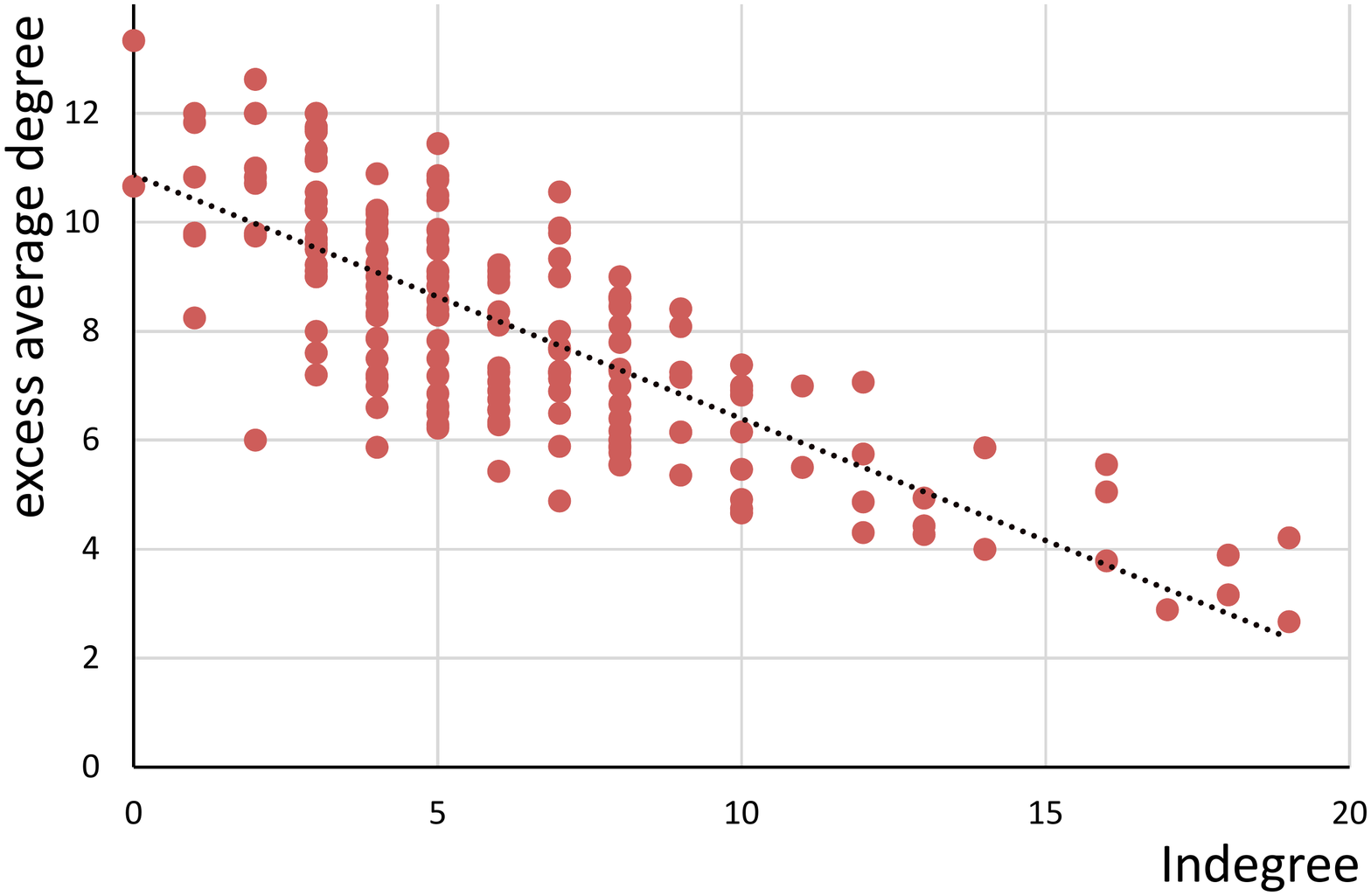}
		\end{minipage}
	}
	\caption{The correlation between indegree and excess average degree for each network.}
	\label{fig:indegree and excess agerage degree} 
\end{figure*}

In Figure ~\ref{Joint probability} we visualized the joint probability distribution of the indegree in all six networks. Note that nodes in all networks with a median indegree (with 8 to 10 nodes pointing to them) can have connections with higher probabilities (but no higher than 0.05, which is relatively very small), as there are more students in the social networks attracting 8 to 10 friends. We can also see this result from the indegree distribution and the excess degree distribution. Additionally, we computed the assortativity coefficient in these six networks, and we noticed that the values are all very low (less than 0.2). This result indicates that there are no obvious correlations between the connections of the indegree of node pairs.

\begin{table}
	\caption{Correlation between three observed features and indegree in six social networks}
	\label{tab:corr}
	\begin{tabular}{ccccccc}
		\toprule
		& Bad News & Friendship & Good News \\
		\midrule
		Attractive & 0.107 & 0.131 & $0.157^{*}$ \\
		Dominant & 0.105 & 0.114 & 0.069 \\
		Trustworthy & $0.152^{*}$ & $0.152^{*}$ & 0.143 \\
		\toprule
		& Social Advice & Academic Advice & Cooperation \\
		\midrule
		Attractive & 0.142 & -0.055 & 0.084 \\
		Dominant & 0.122 & 0.031 & $0.161^{*}$ \\
		Trustworthy & 0.129 & 0.078 & $0.150^{*}$ \\
		\bottomrule
	\end{tabular}
\begin{flushleft}
* $p$<0.05
\end{flushleft}
\end{table}

Figure~\ref{fig:indegree and excess agerage degree} shows the correlation between indegree and excess average degree, which is the average degree of their neighbors. There is an obvious negative correlation in all these six networks. This indicates that students prefer to choose people who are different than themselves.

\begin{figure}[b]
	\includegraphics[height=2in, width=2in]{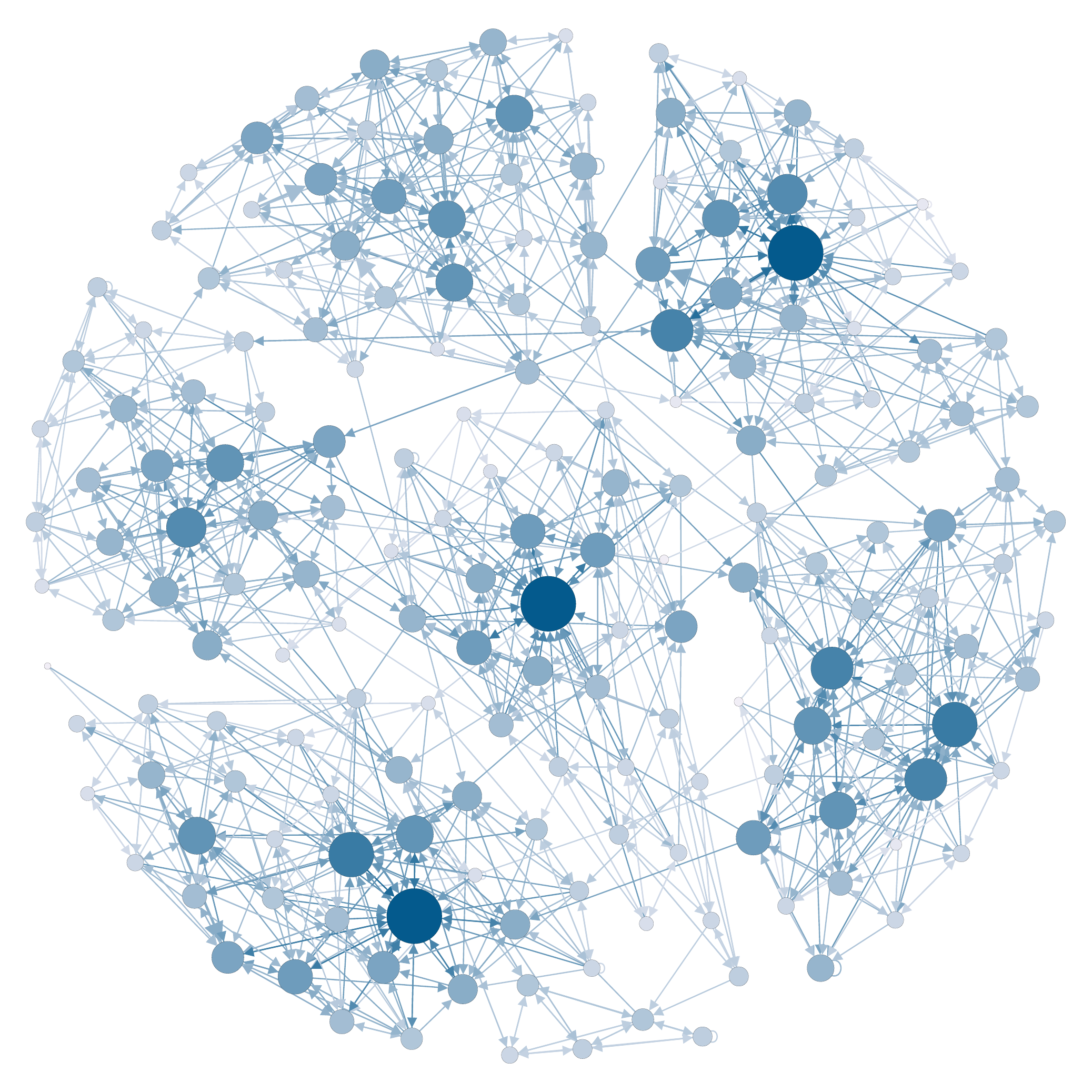}
	\caption{Sketch of the friendship network. The node represents student and the diameter of the nodes in the network indicates the trustworthiness score given by others.}
	\label{fig:ft}
\end{figure}
\begin{figure}
	\includegraphics[height=2in, width=2in]{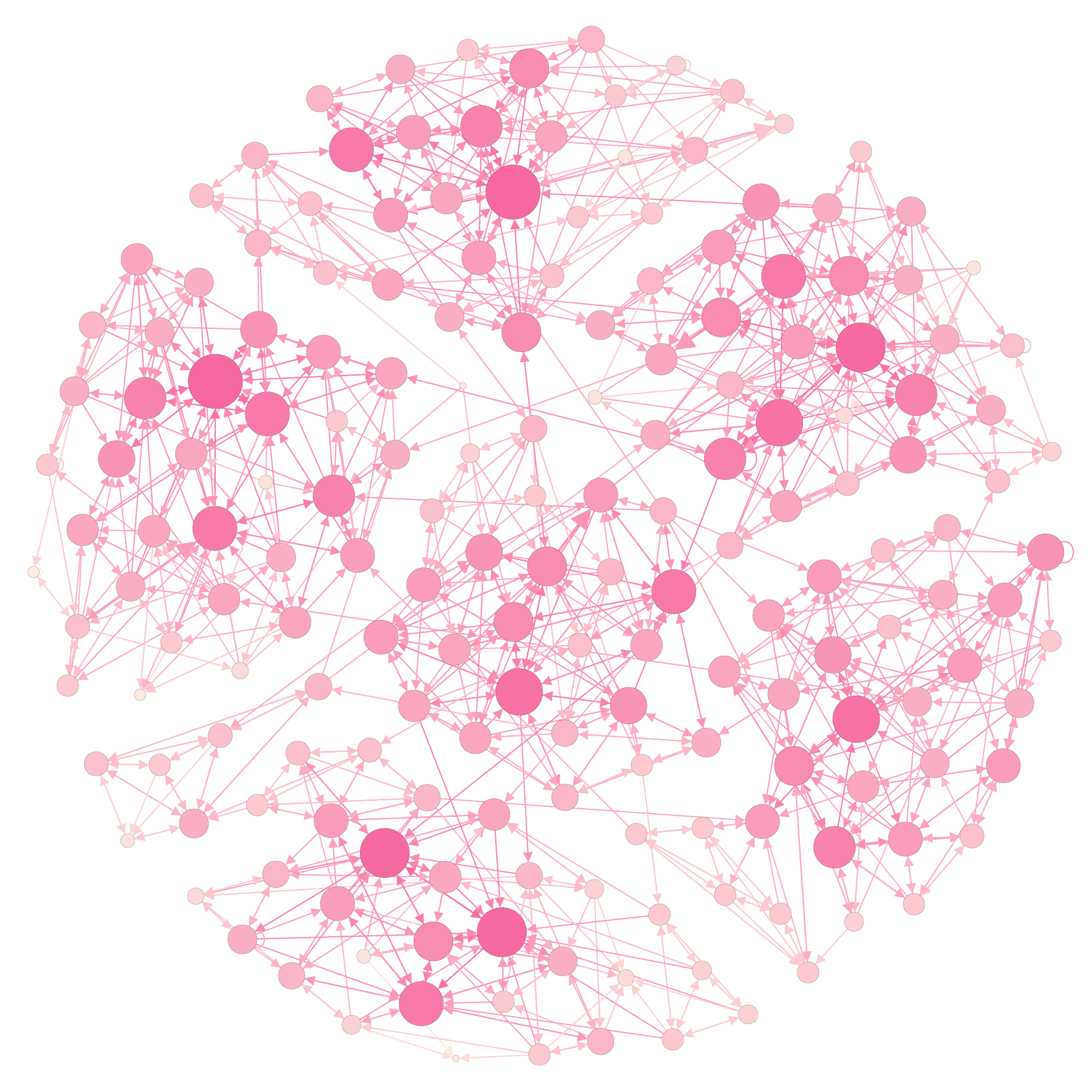}
	\caption{Sketch of the good news network. The node represents student and the diameter of the nodes in the network indicates the attractiveness score given by others.}
	\label{fig:ga}
\end{figure}

\begin{figure*}
	\subfigure[SVC]{
		\begin{minipage}{8.5cm}
			\centering
			\includegraphics[height=5.3cm]{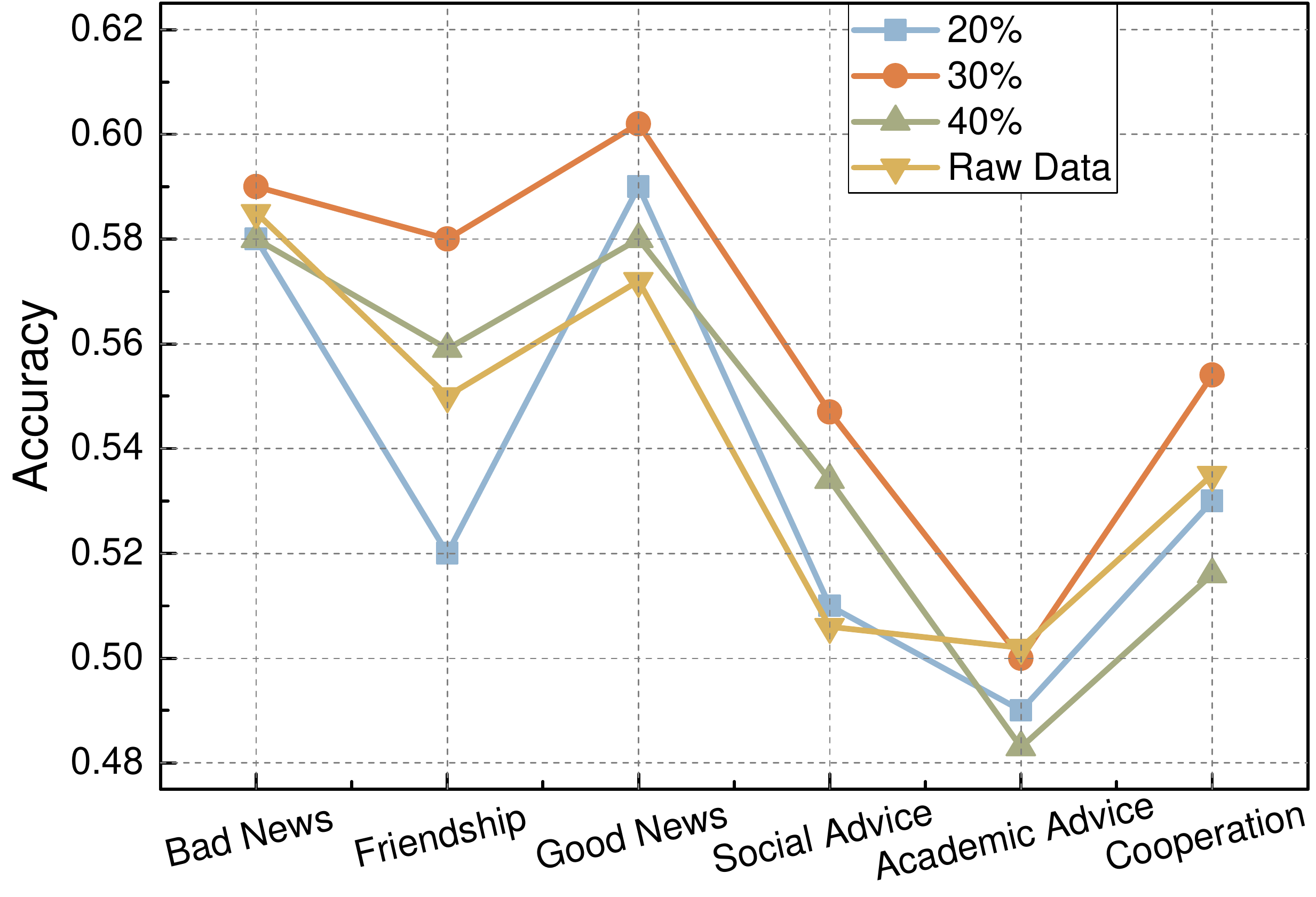}
		\end{minipage}
	}
	\subfigure[LR]{
		\begin{minipage}{8.5cm}
			\centering
			\includegraphics[height=5.3cm]{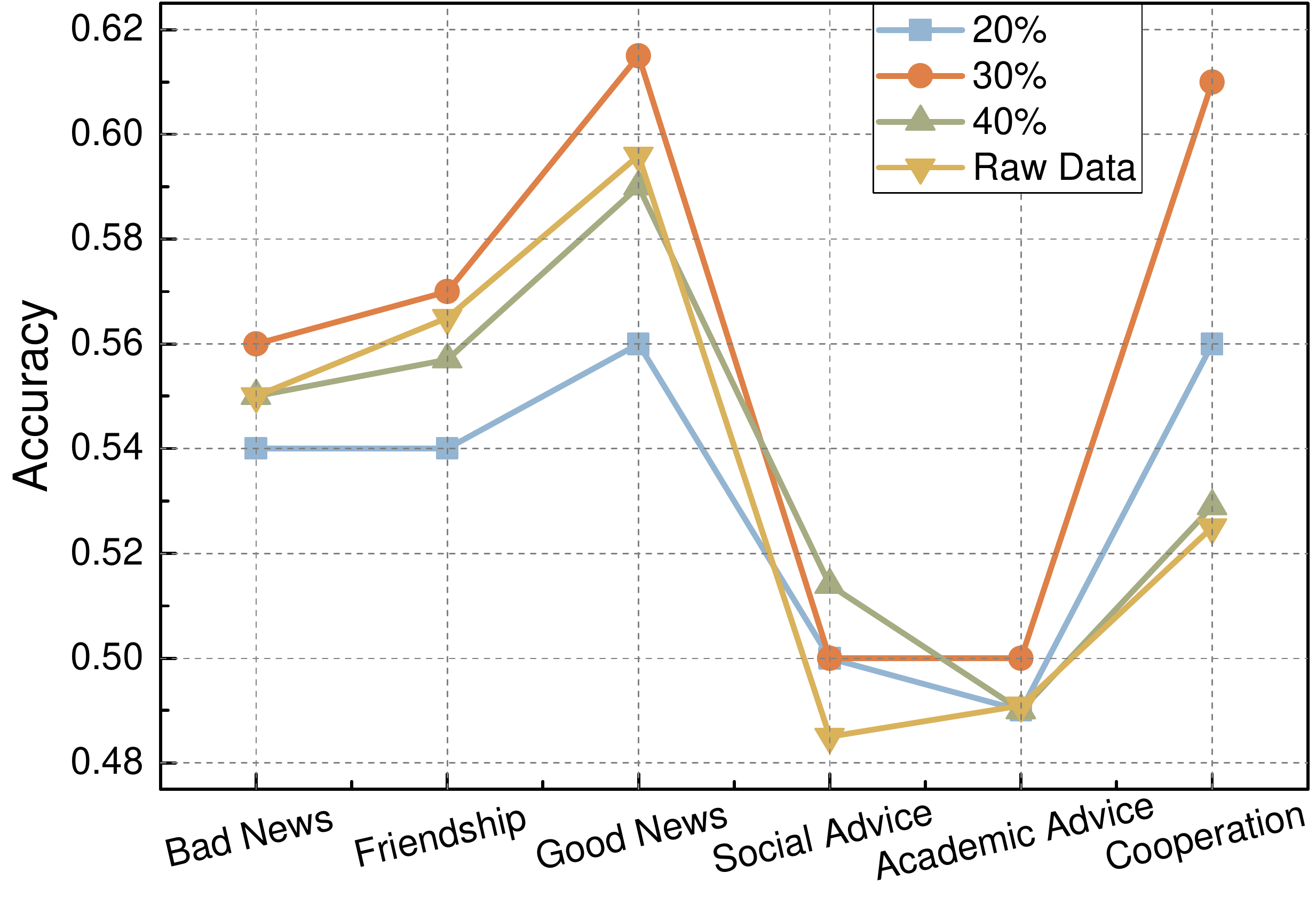}
		\end{minipage}
	}
	\subfigure[NB]{
		\begin{minipage}{8.5cm}
			\centering
			\includegraphics[height=5.3cm]{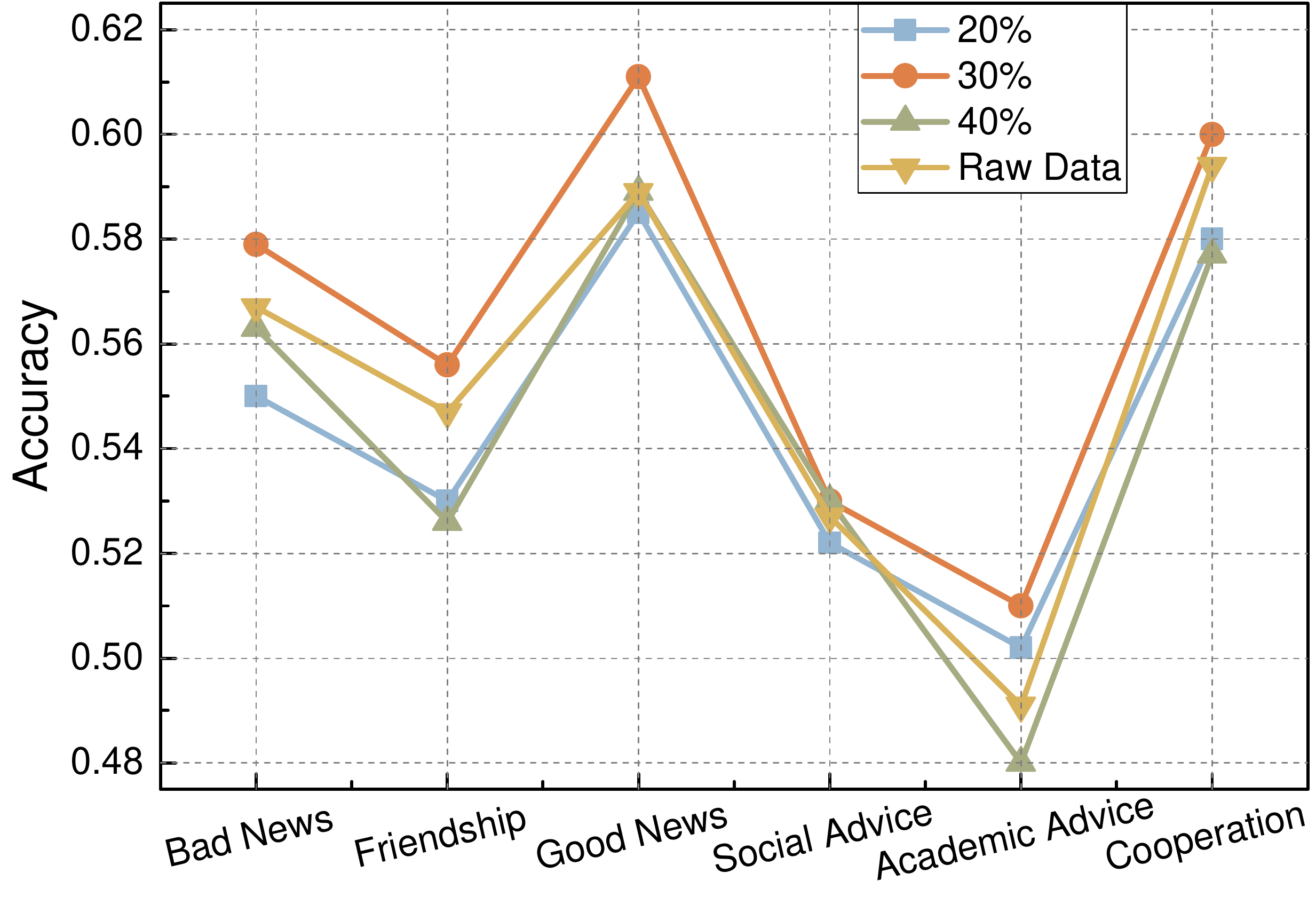}
		\end{minipage}
	}	\subfigure[KNN]{
		\begin{minipage}{8.5cm}
			\centering
			\includegraphics[height=5.3cm]{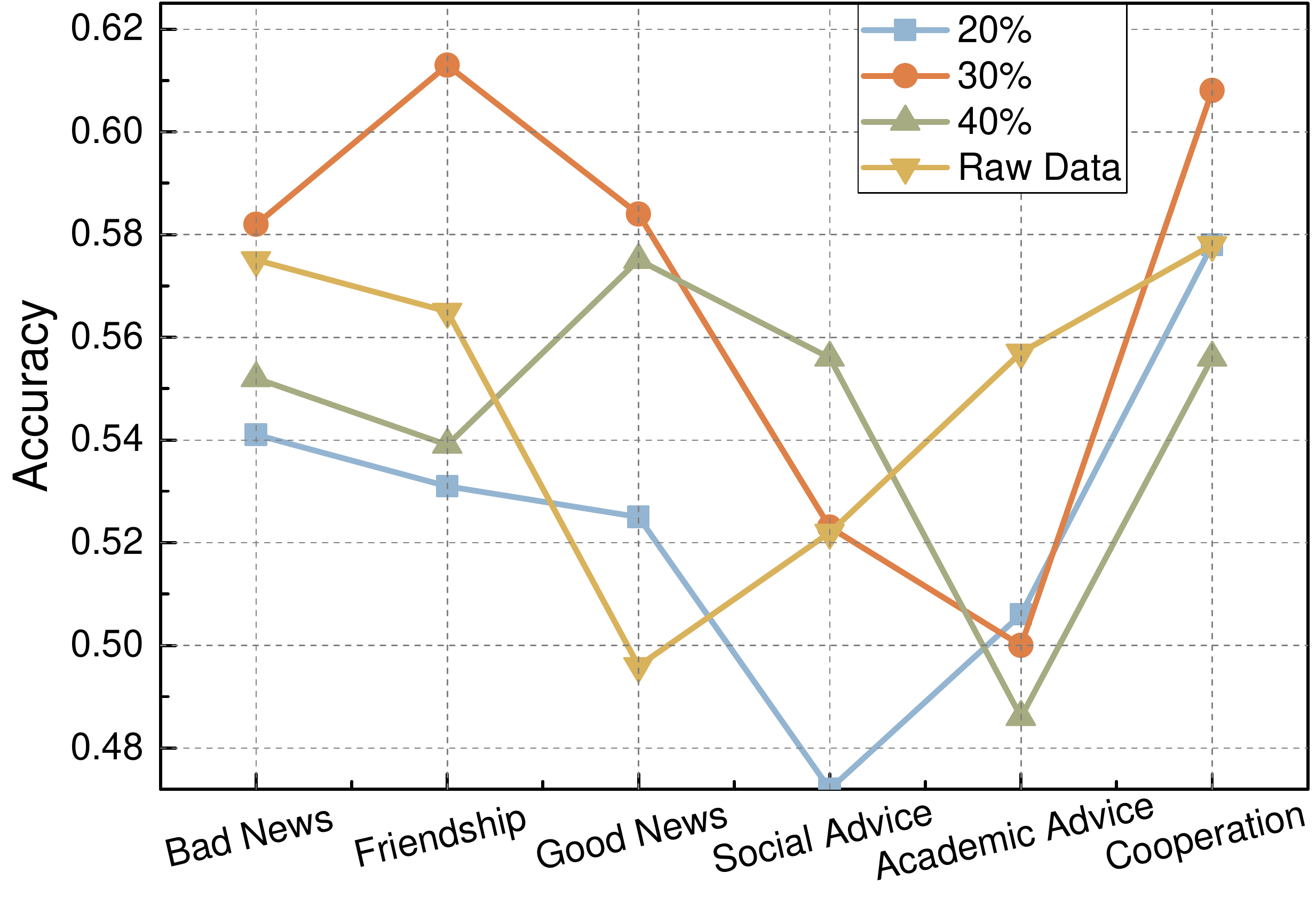}
		\end{minipage}
	}
	\caption{Prediction performance of four classical classification algorithms. Lines of various colours represent the different percentage that we remove the low-entropy data from the raw data.}
	\label{prediction result}
\end{figure*}

\subsection{Correlation Analysis}

To explore the relations between facial appearance and social situation, we performed an analysis on the correlation between the three observed features and the indegrees in six social networks. All findings are shown in Table~\ref{tab:corr}. 

The results show that trustworthiness is related to the indegrees in the bad news and friends networks, as shown in Figure~\ref{fig:ft} (both with a correlation coefficient of 0.152 and p < 0.05). For the choice of people as good friend, trustworthiness is more important than attractiveness and dominance (shown in Figure~\ref{fig:ga}), which confirms a saying by Epicurus: without confidence, there is no friendship. People prefer to share bad news with people who get high scores in trustworthiness, which is consistent with common sense in that people would far rather show their bad side to trustworthy people than those with beautiful facial appearances. This also suggests that part of the role of a friend is to share pain.

When it comes to the good news spreading network, on the contrary, students’ attractiveness shows a significant correlation with indegree (with a correlation coefficient of 0.157 and $p-value < 0.05$). This means that the people with whom one can share bad news are not the same people with whom one can share good news. The purpose of sharing bad news is to share the pain and to get comfort. However, the purpose of sharing good news is not just sharing happiness, but also showing one’s good side to the person with good facial appearance, possibly to reduce the psychological gap caused by facial appearance. \cite{birman2018clinical} 

Compared with the scenarios of choosing friends, sharing good news and sharing bad news, the following three scenarios are relatively rational. In the cooperation network, both dominance and trustworthiness are significant, which indicates that while people want to build a team, the one who shows enough dominant force and looks more trustworthy is more favorable. It is the only time that participants take dominance into account. In psychology, dominance is a measure of a person’s characteristic feelings of control and influence over his/her life circumstances versus feelings of being controlled and influenced by others or events. If attractiveness represents external competitiveness, then dominance represents inherent competitiveness, giving people confidence. In such a rational sense with a clear goal to complete a task, it suggests than students prefer to choose someone with inherent competitiveness over someone with a beautiful appearance. For social advice and academic advice, facial appearance does not have any effect statistically. In these two scenarios, the situation is so complicated, including making decisions on various questions, that it is difficult to make a decision solely according to facial appearance.

In addition, these findings also suggest that the main factors taken into account in various social scenarios differ. Finally, compared with attractive and dominance, trustworthiness is more important for people’s social lives, because three of these six social scenarios involve trust. In other words, being a trustworthy person will give you a good chance of being in a favorable social situation.

\subsection{Prediction}

The significant correlations imply that the three features extracted from facial appearance can predict students’ centrality in different social networks. To explore their predictive power, we carried out an experiment using the score of three aspects from student’s facial appearance to predict their centrality in six different social networks. Excluding some incomplete or error data records, we got 185 samples in the experiment dataset. All three facial appearance quantification scores (attractiveness, dominance, and trustworthy) featured in the six observed networks. As every photo was quantified by 30 participants, we computed the average score from the final points. We performed our prediction task as a binary classification problem, which predicted whether a node could be high degree (with indegree higher than the median of population) or not.

We divided the dataset into two categories by degree of centrality: high and low. We used machine learning to predict the categories of the participants from their scores on the three dimensions. First, we removed 30 participants randomly from the dataset by stratified sampling to use as a test set, and we used the rest of the data as a train set. We performed a fivefold cross-validation on the train set to tune the super parameters of the predictive models, k-nearest neighbors, linear support vector classifier, naive Bayes classifier, and logistic regression, which are long-standing classical binary classification algorithms. 

We calculated the participant entropy of each participant according to the score they gave to each photo. We assumed that there is a positive correlation between participant entropy and the validity of data given by participant. The higher the entropy gets, the more valid the data is. In this study, we removed 20\%, 30\% and 40\% low entropy data respectively to remove the noise existed in data and the performances are shown in Figure~\ref{prediction result}.

Because the proportions of the positive and the negative are equal in the dataset, we used the accuracy value to evaluate the result of prediction. The accuracy value ranges from 0 to 1, with 0.5 being random chance in this condition; therefore, the predictive power is the extent to which the accuracy value exceeds 0.5.

Prediction results are shown in Figure ~\ref{prediction result}. Obviously, it is notable that three features extracted from facial appearance are effective for predicting the centrality of four different networks (friendship network, bad news network, good news network and cooperation network) because all accuracy values exceed 0.5. The prediction result of removing 30\% low-entropy data is the best. The result of removing 20\% data is lower possibly because noise still exist in data, while the result of removing 40\% data is lower possibly because some meaningful data is lost. In addition, accuracies of prediction in the social advice network and the academic advice network are relatively low, which is consistent with the previous findings about their correlation coefficients. 

\begin{comment}
\begin{table}
	\caption{Prediction Results}
	\label{tab:result}
	\begin{tabular}{lc|lc}
		\toprule
		Linear SVC &     & LR &      \\
		\hline
		Network & Accuracy & Network & Accuracy \\
		\hline
		Bad News & 0.70 & Bad News & 0.67 \\
		Good News & 0.73 & Good News & 0.67 \\
		Friendship & 0.67 & Friendship & 0.53 \\
		Social Advice & 0.52 & Social Advice & 0.57 \\
		Academic Advice & 0.53 & Academic Advice & 0.37 \\
		Cooperation & 0.53 & Cooperation & 0.63 \\
		\hline
		Naive Bayes &     & kNN &      \\
		\hline
		Network & Accuracy & Network & Accuracy \\
		\hline
		Bad News & 0.63 & Bad News & 0.57 \\
		Good News & 0.67 & Good News & 0.49 \\
		Friendship & 0.53 & Friendship & 0.56 \\
		Social Advice & 0.40 & Social Advice & 0.51 \\
		Academic Advice & 0.50 & Academic Advice & 0.56 \\
		Cooperation & 0.53 & Cooperation & 0.57 \\
		\bottomrule
	\end{tabular}
\end{table}
\end{comment}

\section{CONCLUSIONS}
In this paper, we collected face perception data from freshmen in college to explore how appearance could impact the individuals’ centrality in different social networks. We selected six questions to capture their social relations in different scenarios, namely a friendship network, a good news network, a bad news network, a social advice network, an academic advice network, and a cooperation network. Experimental results show that perceived facial traits can predict the centrality of different networks. We proposed a PSP framework to discover the effect of facial perception on social situation and the experimental results verify its performance.
Our framework mechanism has a certain universality, which can be applied to the prediction of centrality in different social networks by reducing noise to improve the prediction accuracy.
Our findings suggest that people’s appearance can impact their social situations at an early social stage in which people are not very familiar with each other. To the best of our knowledge, we are the first to explore the influence of facial appearance on centrality in social networks.

Our results also contribute to the psychological knowledge of facial perception, because they are consistent with previous research that facial appearance affects network-based social behaviors. Prior studies only examined the relationship between faces and behaviors in social network settings without using social network analysis and techniques, whose analysis might be limited and might lack depth. This paper sheds light on a novel approach by showing that apparent facial features can be evaluated and predicted by social network analysis and machine learning.  In addition, our findings offer insights on a combination of psychological and social network techniques, and they highlight the function of facial bias in cueing and signaling social traits. 

Our research is limited by the relatively small sample size, because obtaining a larger number of participants would be very challenging. The privacy and confidentiality of human participants was very important in this research, in that participants’ photographs were taken and judged, and their social networks were reported. Accessing a large number of volunteers who are willing to participate such research is a challenge. However, a thorough analysis of social network from a novel, face-based perspective in this paper may remedy the limitations to some extent. Future work could obtaining a larger dataset to facilitate a more in-depth investigation. 

\section*{Acknowledgments}
This work is partially supported by National Natural Science Foundation of China under Grant Nos. 61602079, 61632011, and 61872054, and by Ministry of Education under Grant No. 16YJCZH141.

\bibliographystyle{ACM-Reference-Format}
\balance 
\bibliography{result_and_method}

\end{document}